\documentstyle[11pt]{article}

\def\pagesetup{
\textwidth 6.0in
\textheight 8.5in
\pagestyle{empty}
\topmargin -0.25truein
\oddsidemargin 0.30truein
\evensidemargin 0.30truein
\raggedbottom\parindent=20pt
\baselineskip=14pt}

\pagesetup

\newenvironment{Titlepage}{\parsep=0pt \topsep=0pt}{\relax}

\def\Title#1\endTitle{\begin{center}%
   \baselineskip=16pt 
 \bf #1\\[.5cm]}                    

\def\Author#1#2\endAuthor{\small\it
   {\rm #1}\\[1pt] #2\\[.3cm]}

\def\And{{\rm and }\\[.3cm]}

\def\endAuthors{\end{center}
                \vglue .5cm    
                    }

\newenvironment{Abstract}%
       {\centering\bgroup
          \begin{minipage}{30pc}\small
            \noindent
	    \centerline{\tenrm ABSTRACT}
	    \parindent=0pt}%
         {\end{minipage}\egroup\par
         \normalsize}

\begin{document}

\begin{Titlepage}
\vskip -0.5truein
\vbox{
\rightline{\small McGill/94--40}
\rightline{\small IASSNS-HEP-94/65}
\rightline{\small  hep--th/9409177}
}

\medskip
\Title
STRINGY TWISTS OF THE TAUB--NUT
METRIC\footnote{Talk presented by RCM
at the Seventh Marcel Grossmann Meeting on General Relativity,
Stanford University,
July 24th---30th, 1994.}
\endTitle
\Author{CLIFFORD V. JOHNSON}
School of Natural Sciences, Institute for Advanced Study,
Princeton, NJ 08540 USA
\endAuthor
\And
\Author{ROBERT C. MYERS}
Department of Physics, McGill University, Montr\'eal, Qu\'ebec,
H3A 2T8 Canada
\endAuthor
\endAuthors

\begin{Abstract}
In the low-energy limit, string theory has two remarkable 
symmetries, $O(d,d+p)$ and $SL(2,{{\rm I}\!{\rm R}})$. We illustrate
the use of these transformations as techniques
to generate new solutions by applying them to the Taub--NUT metric.
\end{Abstract}

\end{Titlepage}

\vglue 0.4cm
Since it is hoped that string theory can provide a consistent theory of
quantum
gravity, there has been a great deal of interest in studying non-trivial
gravitational solutions of the effective string equations of
motion. 
These equations arise from a low-energy effective field-theory
action, which reproduces the interactions of the massless particles
appearing in the string theory.
We write the four-dimensional effective action
for the heterotic string as
\begin{equation}
I=
\int d^4x\,\sqrt{-G}e^{-\Phi}\,\left(R(G)+(\nabla\Phi)^2-{1\over12}H^2
-{1\over8}F^2\right)\ \ . \label{actg}
\end{equation}
The fields considered here are the metric $G_{\mu\nu}$, the dilaton $\Phi$,
the axion $B_{\mu\nu}=-B_{\nu\mu}$ (which appears in the
three-form $H$), and an Abelian gauge field $A_\mu$.
The full heterotic string theory would include many other
massless fields ({\it e.g.,} more gauge fields, fermions, moduli fields,
etcetera),
which may consistently be set to zero. Eq.~(\ref{actg}) also omits an
infinite series of higher-derivative interactions, whose contributions
will be negligible for fields varying slowly on the Planck scale.
The axion field strength has components
$H_{\mu\nu\rho}=\partial_\mu B_{\nu\rho}+\partial_\nu
B_{\rho\mu}+\partial_\rho B_{\mu\nu}-\omega(A)_{\mu\nu\rho},$
where $\omega(A)_{\mu\nu\rho}={1\over4}(A_\mu F_{\nu\rho}+A_\nu
F_{\rho\mu}+A_\rho F_{\mu\nu})$
is the Chern-Simons three-form for the gauge
field. The gauge field
strength is $F_{\mu\nu}=\partial_\mu A_\nu-\partial_\nu A_\mu$.

One set of low-energy solutions can be found by setting
$\Phi=0$, $B_{\mu\nu}=0$ and $A_\mu=0,$ in which case the equations
of motion reduce to $R_{\mu\nu}=0$. Thus any metric which is a
solution of the vacuum Einstein equations provides a solution of the
effective string theory. One such example is the
Taub--NUT solution\cite{TaubNUT}:
\begin{equation}
ds^2 =-f_1
(dt+2l\cos\theta d\phi)^2
+f_1^{-1}dr^2+(r^2+l^2)(d\theta^2+\sin^2\theta d\phi^2)
 \label{taubnut}
\end{equation}
where
\[
f_1\equiv 1-2{Mr+l^2\over r^2+l^2}\ .
\]
One may wish to generalize this solution by introducing non-trivial
electromagnetic fields. A charged Taub--NUT solution of the
Einstein-Maxwell equations is known\cite{brill},
but it will not provide a solution
of the string equations because of the non-trivial dilaton and axion
couplings to the gauge field
in the effective action (\ref{actg}). These couplings, though,
lead to some remarkable symmetries in the low-energy theory,
which allowed us to produce a new dyonic Taub--NUT solution.\cite{cvj2}

The first of these symmetries applies to background fields
with Killing symmetries. Given a class of solutions
with $d$ commuting Killing symmetries and $p$ Abelian gauge fields, there
exists an $O(d,d+p)$ symmetry of the action\cite{odd}.
Many of these transformations actually produce pure gauge and coordinate
transformations, but the remaining transformations generate new solutions.
Using the time-translation symmetry of Eq.~(\ref{taubnut}),
we applied an $O(1,2)$ boost to produce a new solution
with both electric and magnetic fields.
This transformation introduces a single new parameter, and so
the resulting electric and magnetic charges were not independent.

A generalized dyon with independent charges can be constructed
using the second symmetry\cite{frank}, $SL(2,{{\rm I}\!{\rm R}})$.
These transformations are
a string version of the conventional
electromagnetic duality rotations,
which trade electric and magnetic fields. The
$SL(2,{{\rm I}\!{\rm R}})$ transformations also act non-trivially
on the axion and dilaton fields.

After all of the transformations, the final metric is:
\begin{equation}
\label{newsigma}
d{s}^2={f_2{}^2+f_3{}^2\over1+y^2}\left[
-{f_1\over f_2{}^2}d\xi^2
+{dr^2\over f_1}
+(r^2+l^2)(d\theta^2+\sin^2\theta\,d\phi^2)\right],
\end{equation}
where $d\xi=dt+(x+1)l\cos\theta\,d\phi$,
\[
f_2\equiv 1+(x-1){Mr+l^2\over r^2+l^2}\qquad{\rm and}\qquad
f_3\equiv (x-1)l{r-M\over r^2+l^2}-y\ \ .
\]
The additional non--vanishing fields are:
\begin{eqnarray}
\label{morepot}
{A}_t&=&2
\,z\,f_4\qquad\qquad\qquad\qquad\qquad
{\rm e}^{\textstyle -{\Phi}}={(y^2+1)\,f_2\over f_2{}^2+f_3{}^2}
\nonumber\\*
{A}_\phi&=&2
\,z\,\cos\theta \left(M-ly+
l(x+1)f_4\right)\nonumber\\*
{B}_{\phi t}&=&{x-1\over y^2+1}{\cos\theta\over (r^2+l^2)\,f_2}
\left((l+2My-ly^2)(r-M)^2\right.\nonumber\\*
&&\qquad\qquad\qquad\left.+(lx-2My+ly^2+(x+1)yr)(M^2+l^2)\right)
\end{eqnarray}
where $f_4\equiv[l(r-M)+y(Mr+l^2)]/[(r^2+l^2)\,f_2]$
and $z^2\equiv(x^2-1)/(y^2+1)$.

The mass and NUT parameters, $M$ and $l$, of the original solution
are supplemented by two new independent parameters:
$x$ arises in the $O(1,2)$ boost
with $x^2\ge1$, and $y$
arises from $SL(2,{{\rm I}\!{\rm R}})$ and may take any real value.
By examining geodesics in the asymptotic region, we determine the
mass of the new solution (\ref{newsigma}) to be
${\mu}={x+1\over2}M.$
Similarly, it is
straightforward to calculate the electric and magnetic charges:
${Q}_E=2z(l+yM),$ ${Q}_M=-2z(M-yl).$
An interesting property of the Taub--NUT solution (\ref{taubnut}) is
that the time coordinate must be identified with period $8\pi\,|l|$
to avoid conical singularities at $\theta=0$ and $\pi$.
Surfaces of constant radius then have the topology of a three-sphere,
in which there is a Hopf fibration of the $S^1$ of time over the spatial
$S^2$. The new dyonic solutions (\ref{newsigma}) retain this property
(for $x\ne-1$), but the time period is now $4\pi|l(x+1)|$. While
the original solution contains no curvature singularities, the dyonic
solutions have singularities when $l^2\le{(x-1)^2\over4x}M^2$ at the
radii where $f_2=0$.

A special case of these solutions is the extremal limit,
which is produced by taking $M,l\rightarrow 0$ and
$x\rightarrow\infty$, while keeping $m=xM$ and $\lambda=xl$ fixed.
A metric describing the throat region is produced by simultaneously
scaling the radial coordinate.\cite{cvj2} One can then show that
the solution with $y=0$ corresponds to the string-theory background
described by an exact conformal field theory.\cite{cvj1}

The most remarkable feature of these string constructions is that
extremely complex solutions, such as in Eqs.~(\ref{newsigma}) and
(\ref{morepot}), are produced by algebraic calculations.
These dyonic Taub--NUT solutions were also constructed independently
in Refs.~[7,8] 
A generalization
of these solutions including angular momentum was also found
in Ref.~[9]. 

\section*{Acknowledgements}

CVJ was supported by a EPSRC (UK) postdoctoral fellowship. RCM was
supported by NSERC of Canada, and Fonds FCAR du Qu\'ebec.

\end{document}